\documentclass[twocolumn,preprintnumbers,amssymb,amsmath,nofootinbib]{revtex4}
\pdfoutput=1
\usepackage{graphicx}
\usepackage{amsfonts}
\usepackage{lipsum}
\usepackage{amssymb}
\usepackage{amsthm}
\usepackage{amsmath}
\usepackage{hyperref}
\usepackage{slashed} 
\usepackage{color}
\def\eq#1{{eq.~(\ref{#1})}}

\def\Tr{\mbox{Tr}\,}

\def\di{\mbox{d}}

\def\gtap{\ \raisebox{-.4ex}{\rlap{$\sim$}} \raisebox{.4ex}{$>$}\ }

\definecolor{oucrimsonred}{rgb}{0.6, 0.0, 0.0}
\definecolor{persianblue}{rgb}{0.11, 0.22, 0.73}
\definecolor{forestgreen}{rgb}{0.13,0.35,0.13}
 \hypersetup{colorlinks, citecolor=oucrimsonred, linkcolor=persianblue, urlcolor=oucrimsonred}
\def\hhref#1{\href{http://arxiv.org/abs/#1}{#1}} 
\newcommand{\be}{\begin{equation}}
\newcommand{\ee}{\end{equation}}
\newcommand{\bea}{\begin{eqnarray}}
\newcommand{\eea}{\end{eqnarray}}
\newcommand{\nn}{\nonumber}

\begin{document}
\title[]{Constraints on  top quark non-standard interactions\\
 from Higgs and $t \bar t$ production  cross sections
}
\date{\today}
\author{D. Barducci$^{\dag}$}
\author{M. Fabbrichesi$^{\ddag}$}
\author{A. Tonero$^{\circ}$}
\affiliation{$^{\dag}$SISSA and INFN, Sezione di Trieste, via Bonomea 265, 34136 Trieste, Italy}
\affiliation{$^{\ddag}$INFN, Sezione di Trieste, Via Valerio 2, 34127, Trieste, Italy}
\affiliation{$^{\circ}$ UNIFAL-MG, Rodovia Jos\'e Aur\'elio Vilela 11999, 37715-400 Po\c{c}os de Caldas, MG, Brazil }
\begin{abstract}
\noindent  We identify the differential cross sections for $t\bar t$  production
and  the total cross section for Higgs production through gluon fusion as the processes in which the two  effective operators  describing the leading non-standard interactions  of the top quark with the gluon  can be disentangled and studied in an independent fashion. 
Current data on the Higgs production and the $ \di \sigma/\di {p^t_T}$ differential cross section   provide  limits comparable, but not  more stringent,  than those from the total $t\bar t$ cross sections measurements at the LHC and Tevatron, where however the two operators enter on the same footing and can only be constrained together. 
We conclude by stating  the (modest) reduction  in the uncertainties necessary to provide  more stringent limits by means of  the Higgs production and $t\bar t$ differential cross section observables  at the LHC with the future luminosity of 300 and 3000~fb$^{-1}$.

\end{abstract}

\maketitle
\section{Introduction}
\label{sec:mot}

The top quark is the heaviest among the standard model (SM) quarks and is therefore the best candidate to be studied for any departure from  particle point-like behavior. Such a departure would  point to physics beyond the SM, possibly related to the dynamics behind the electro-weak symmetry breaking (EWSB) mechanism. 

The Large Hadron Collider (LHC) is expected to produce by the end of its run-3, with a collected integrated luminosity of $300\;$fb$^{-1}$, roughly $2\times 10^8$ top quarks pairs,  effectively acting as a \emph{top factory} and thus providing
the possibility of scrutinising the top quark intrinsic properties with an unprecedented precision. 
Moreover, the top quark  enters in the dominant Higgs production mechanism at the LHC, the production via gluon fusion, which is also expected to be measured with high accuracy by the end of the LHC program.

The study of the properties of the top quark has been performed both in terms of anomalous  couplings~\cite{Grzadkowski:2003tf,Lillie:2007hd,Choudhury:2009wd,Hioki:2013hva,Kamenik:2011dk,Biswal:2012dr,Rindani:2015vya} and  of SM higher dimensional effective 
operators~\cite{Kumar:2009vs,Zhang:2010dr,Zhang:2012cd,Degrande:2010kt,Englert:2012by,Degrande:2012gr,Cirigliano:2016nyn,Englert:2016aei}, often with an overlap between the two approaches. 
While the anomalous-coupling approach has the advantage of a more direct physical interpretation and the lower number of parameters, the effective lagrangian framework provides a more general and unbiased view, 
based on the possibility of performing global fits on a larger number of operators affecting various processes, see {\emph{e.g}}.~\cite{Buckley:2015nca,Englert:2015hrx}.

In this work, motivated by the fact that strong interactions  dominate  $t\bar t$ production at the LHC, we follow the anomalous coupling approach, by studying the top-quark hypothetical structure only by means of its interaction to gluons. We parametrize it in terms of the following $SU(3)_C \times U(1)_{em}$
effective operators
\be
{\cal O}_1 = \frac{C_1}{\Lambda^2}\, \bar t \gamma^\mu T^a t \, D^\nu G^a_{\mu\nu} \label{q1} 
\ee
\be
{\cal O}_2 = \frac{C_2}{\Lambda^2}\, v \, \bar t \sigma^{\mu\nu} T^a t \, G^a_{\mu\nu}  \label{q2}\, ,
\ee
where $T^a = \lambda^a/2$ are the $SU(3)_{C}$ generators, $[T^a,T^b]=i f_{abc}T^c$ and Tr$[\lambda^a \lambda^b]=2 \delta^{ab}$, 
$D^\nu=\partial^\nu-i g_s G^{\nu,a}T^a$ and $G^a_{\mu\nu}=\partial_\mu G^a_\nu-\partial_\nu G^a_\mu + g_s f^{abc}G_\mu^b G_\nu^c$ are the $SU(3)_C$ covariant derivative and the field strength tensor respectively and $\sigma^{\mu\nu} = i/2 [\gamma^\mu,\gamma^\nu]$.
These two effective operators can also be seen as the leading terms coming from the Taylor expansion of the strong version of the  Dirac and Pauli form factors in the top gluon interaction~\cite{Fabbrichesi:2013bca}, thus making perhaps more evident the relationship with the study of the internal structure of the top quark. This point will be discussed in Section~\ref{sec:form_factors}.
The  vacuum expectation value $v=174$ GeV in Eq.~\eqref{q2} is a reminder of the presence of the Higgs boson in the $SU(2)_L$ invariant operator before EWSB. This will induce further interactions affecting Higgs phenomenology which we will discuss in Sec.~\ref{sec:higgs}.

The effective operators of Eq.~\eqref{q1} and Eq.~\eqref{q2} affect both $t\bar t$ and Higgs production processes, which can then be used to constrain the corresponding Wilson coefficients. 
The relation between the operator $\mathcal O_1$ of Eq.~\eqref{q1} and the four-fermions operators in the Warsaw basis~\cite{Grzadkowski:2010es} is given in Appendix~\ref{appendix}.
Because of its space-time structure, the three point function arising from the operator ${\cal O}_1$ vanishes when coupled to on-shell gluons, and thus does not affect the dominant Higgs production mechanism at the LHC, the one via gluon fusion, which can then be used to constrain the size of the operator ${\cal O}_2$ independently of  ${\cal O}_1$. On the other hand, even though  the operator ${\cal O}_2$ enters both processes, its contribution only marginally modifies the shape of the top quark pair invariant mass and transverse momentum distributions~\cite{Franzosi:2015osa}. 
Modifications are present in the high energy regime when quadratic terms in $C_2$ are retained. Therefore for small values of the Wilson coefficient negligible departures with respect to the SM predictions are expected. 
In order words, the shapes of the normalized $1/\sigma\;\di\sigma/\di{m_{t \bar t}}$ and $1/\sigma\;\di\sigma/\di{p_{T}^t}$ distributions are essentially unaffected by the presence of the ${\cal O}_2$ operator. 

We  thus conclude that the combined  study of the inclusive Higgs production and of the differential cross sections for $t\bar t$ production could offer two observables constraining the operators ${\cal O}_1$ and ${\cal O}_2$ independently of each other, thus providing, in principle, more stringent limits than those we can obtain from other processes, like the total cross section, where the simultaneous presence of both operators requires some marginalization in order to set the constraints. 

As we will show, the use of these independent observables to set more stringent  limits is only possible  if the uncertainties in the differential cross section measurements can be reduced, especially in the high momentum-transfer region. While this is expected to happen as more data will be collected, it is not the case yet for those currently available. For this reason we still use the total $t\bar t$ production cross section---where both ${\cal O}_1$ and ${\cal O}_2$ enter--- to set the strongest limits available today. 
We  then identify the expected reduction in uncertainty necessary  to have the $t\bar t$ differential cross section and Higgs production process to set the most stringent limit on the operator ${\cal O}_1$ independently of ${\cal O}_2$ at the LHC with the future luminosity of 300 and 3000 fb$^{-1}$.

\subsection{Form factors and gauge invariance}
\label{sec:form_factors}

The physical interpretation of the contribution of the operators in Eq.~\eqref{q1} and Eq.~\eqref{q2} to cross sections measurements is in terms of a departure from the point-like behavior of the top quark. From this point of view,
as already mentioned in the previous section, these  operators  can be seen as the leading terms coming from the Taylor expansion of the strong version of the electromagnetic form factors. 

In order to explain this point, it is useful to first recall how nucleon electromagnetic form-factors are defined. They are usually introduced through an effective parametrization of the nucleon-photon vertex $\Gamma_\mu$ which in momentum space reads as follows:
\be \label{emff}
\Gamma_\mu(q,k)=e\, \gamma_\mu F_1(q^2)+i e\,\frac{\sigma_{\mu\nu}}{2M}q^\nu F_2(q^2)\,,
\ee
where $q$ is the photon momentum and $F_1$ and $F_2$ are. respectively, the Dirac and Pauli form factors, with $F_1(0)=1$ and $F_2(0)=\kappa$. This parametrization of the vertex respects electromagnetic gauge invariance when considering on-shell external nucleons.

In the case of strong interactions, where the underlying $SU(3)_C$ symmetry is non-abelian, a parametrization similar to that of Eq.~\eqref{emff} would violate gauge invariance. Therefore, form factors that respect gauge invariance have to be introduced by considering, in addition to the covariant kinetic term, the following operators:
\be \small 
\label{strff}
\bar \psi \left[  \frac{C_1}{\Lambda^2}\gamma^\mu f_1\left (\frac{D^2}{\Lambda^2} \right) D^\nu G_{\mu\nu}+\frac{C_2}{\Lambda}\sigma^{\mu\nu}f_2 \left( \frac{D^2}{\Lambda^2} \right)  G_{\mu\nu}  \right]\psi\,,
\ee
where $D^2 =D_\mu D^\mu$. The functions $f_1$ and $f_2$ are the strong analogous  of the Dirac and Pauli form factors. These form factors are assumed to admit a Taylor expansion. 
The leading terms of the expansion is what we consider in our study and are represented by the operators introduced in Eq.~\eqref{q1} and Eq.~\eqref{q2}. 

While, in the case of electromagnetic interactions, form factors can be introduced in a way that their presence affects just the interaction vertex between one single photon and the fermion, in the case of strong interactions, gauge invariance requires that form factors affect also interaction vertices between the fermion and a multiple number of gluons. This can be seen by expanding the functions $f_1$ and $f_2$ in Eq.~\eqref{strff} and substituting the explicit expression of the covariant derivative. 

\subsection{The fine print}
\label{sec:fine_print}

The reliability of the perturbative expansion of the effective theory depends on the relative size of the higher order operators with respect to those we retain in the cross section. This size is controlled by the energy of the process, the energy scale of the effective theory  and the estimated strength of  couplings. Concerning the leading corrections to the SM result, the size of which is controlled by $g_{SM}$,  and indicating with  $\bar E$  the energy probed in the process, we have  terms
\be
 O\left( \frac{g_{SM} C^{(6)}\bar{E}^2}{\Lambda^2}\right)  \, , \label{inter}
\ee
which arise from the interference  between the SM amplitude and the leading dimension-six operators,   terms
\be
O\left( \frac{C^{(6)} \bar{E}^2}{\Lambda^2}\right)^2  \label{square} \, ,
\ee
which come from the square (or the double insertion) of the same dimension-six operators, and terms
\be
 O\left( \frac{g_{SM} C^{(8)}\bar{E}^4}{\Lambda^4}\right) \label{8} \, ,
\ee
which originate from the interference between the SM amplitude  and the dimension-eight operators.  

The terms in Eq.~\eqref{8} are formally comparable to those  in Eq.~\eqref{square}.
Without any assumption about the strength of the interactions behind the effective operators, it is not possible to decide whether the terms in Eq.~\eqref{8} should be included or can be safely neglected. To make such an assumption manifest we can re-write the coefficients $C^{(6)}$ and $C^{(8)}$ as $g_\star \tilde C^{(6)}$ and $g_\star \tilde C^{(8)}$, where $g_\star$ indicates the strength of these interactions. Accordingly, the condition for   the terms in Eq.~\eqref{8} to be smaller than those in Eq.~\eqref{square} is simply
\be
g_\star > g_{SM} \, . \label{gg}
\ee
 In our study, we look into departures from point-like behavior of the top quark. It is then reasonable to assume that such physics originates in interactions that are at least stronger than those of the standard model. This assumption makes  the condition in Eq.~\eqref{gg}  satisfied.
 This argument must be taken with a grain of salt: it is an assumption that $C^{(8)} = g_\star \tilde C^{(8)}$,  rather than higher powers of $g_\star$, and it is another assumption that the numerical coefficients are sufficiently small for making dimensional analysis valid. 

When the terms in Eq.~\eqref{inter} are larger than the SM result itself, it is necessary to  include also those in Eq.~\eqref{square} in order to make a likelihood test well defined (this point was already made in \cite{Contino:2016jqw}). The reason is that otherwise the observable to be estimated could  be negative for negative values of the coefficient $C^{(6)}$. This can happen if the energy $\bar E$ in the process is large enough to overcome in Eq.~\eqref{inter} the suppression from $O(g_{SM} C^{(6)}/\Lambda^2)$.  This is the case in our estimate of the total and differential cross sections  because $\bar E=m_{t\bar t}$ (where $m_{t\bar t}$ is the invariant mass of the system of top quark pairs) can become large enough. We therefore must include the terms in \eq{square}.  On the other hand, the cross section for the Higgs boson production is safe because $\bar E = m_H$  (where $m_H$ is the mass of the Higgs boson) and we can keep only terms of the type  of Eq.~\eqref{inter}.

Another comment  is in order. 
Next-to-leading  order (NLO) corrections to the processes under consideration are crucial in order to match the theoretical predictions with the experimental measurements. It is therefore in principle  necessary to evaluate all amplitudes at least to this order, both for the SM and in the case of the presence of the operators ${\cal O}_1$ and ${\cal O}_2$ of Eq.~\eqref{q1} and Eq.~\eqref{q2}.
It has however been recently shown~\cite{Franzosi:2015osa} that these corrections, at least for what concerns the operator ${\cal O}_2$, only affect the cross section by an overall $k$-factor which is equal for the SM and for the SM augmented by the operator ${\cal O}_2$. This holds true both for the total cross section, as well for the differential ones, where the $k$-factors are now approximately equal to each other bin per bin. Pending a formal proof, we will assume that the same holds true also for the operator  ${\cal O}_1$. For this reason, we perform our calculation at the leading order (LO).

\section{Top pair production cross section measurements}
\label{sec:top}

In order to calculate total and differential event rates for the $t\bar t$ process, we implement the operators of Eq.~\eqref{q1} and Eq.~\eqref{q2}  in the {\tt UFO}~\cite{Degrande:2011ua} format through the {\tt Feynrules}~\cite{Alloul:2013bka} package and use {\tt MadGraph5\_aMC@NLO}~\cite{Alwall:2014hca} as event generator. We then analyse the generated event via the {\tt MadAnalysis5}~\cite{Conte:2012fm} package. 

We perform our calculation at the leading order  and in comparing our results with the $t\bar t$ rates (both total and differential) we assume that the central value of the experimental measurement corresponds to the SM predicted cross section, computed by fixing $C_1=C_2=0$ in our numerical calculation. In other words, we are computing {\emph{expected limits}} on the two Wilson coefficients, as it is usually done when calculating limits for projected measurements. In the case of actual data, we are assuming that the mismatch between the measured values and the SM predictions, when folded with the relevant $k$ factors, are due to statistical fluctuation that we ignore. 

\subsection{Limits from the total  cross section}
\label{sec:tt_total}

The contribution of the two operators in Eq.~\eqref{q1} and Eq.~\eqref{q2} to  the total cross section for $t \bar t$  production has  been previously estimated and limits on their size obtained. The most recent analysis of the two operators taken by themselves can be found in~\cite{Fabbrichesi:2013bca}, 
while one considering the full set of operators affecting top quark phenomenology has been presented in~\cite{Buckley:2015nca} by the use of the dedicate package {\tt TopFitter}.

\begin{figure}[ht!]
\begin{center}
\includegraphics[width=0.22\textwidth]{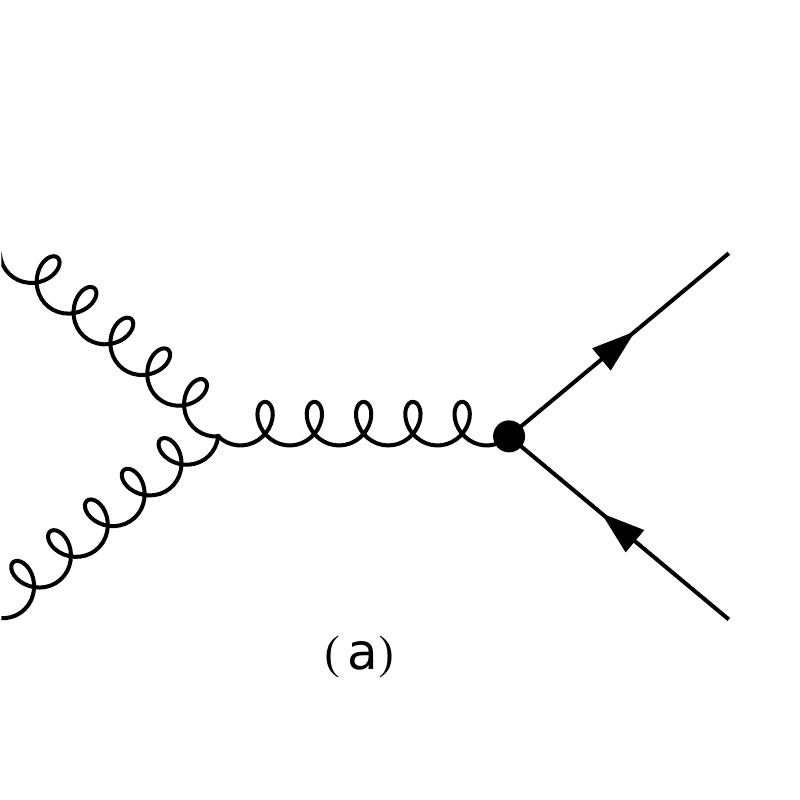}
\includegraphics[width=0.22\textwidth]{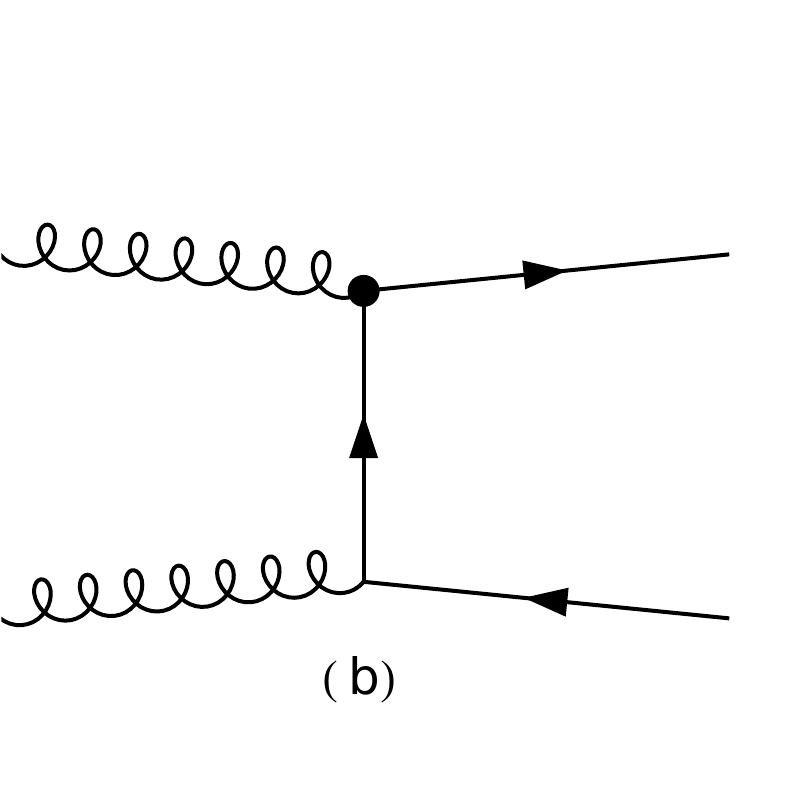}\\
\includegraphics[width=0.22\textwidth]{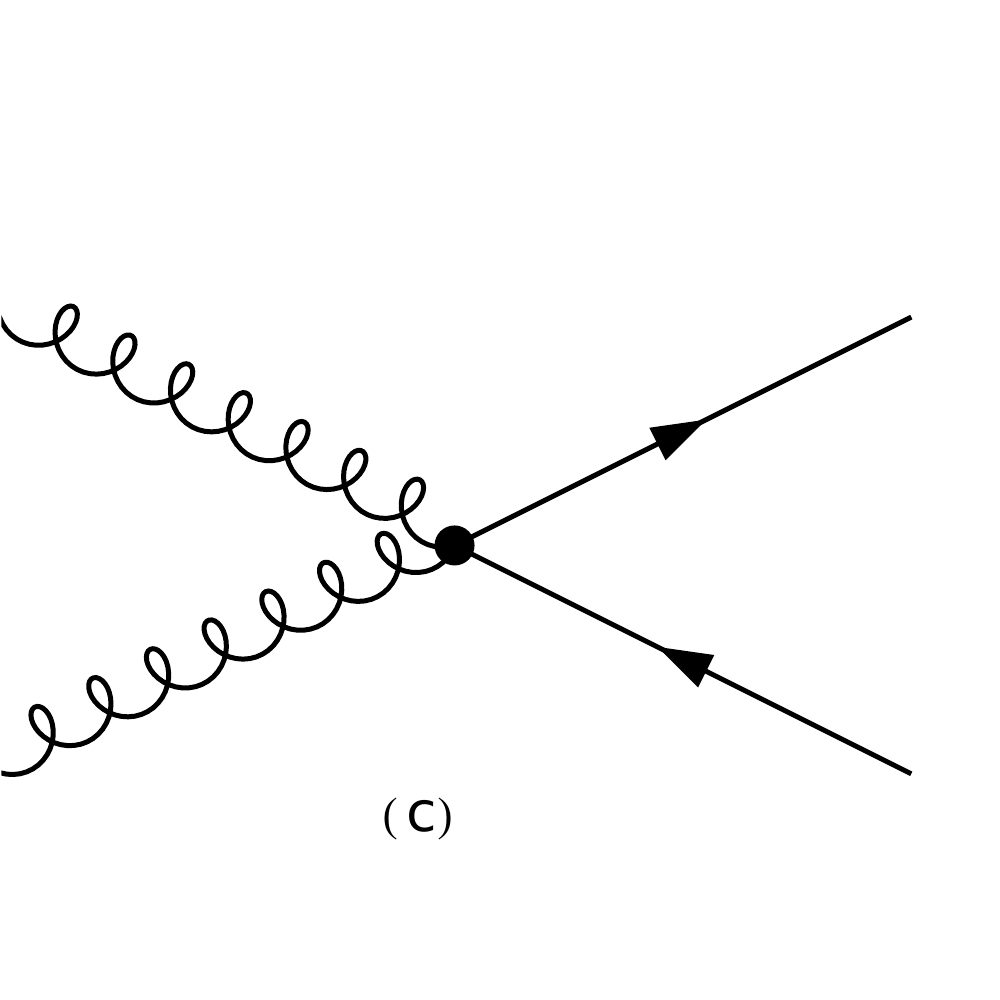}
\includegraphics[width=0.22\textwidth]{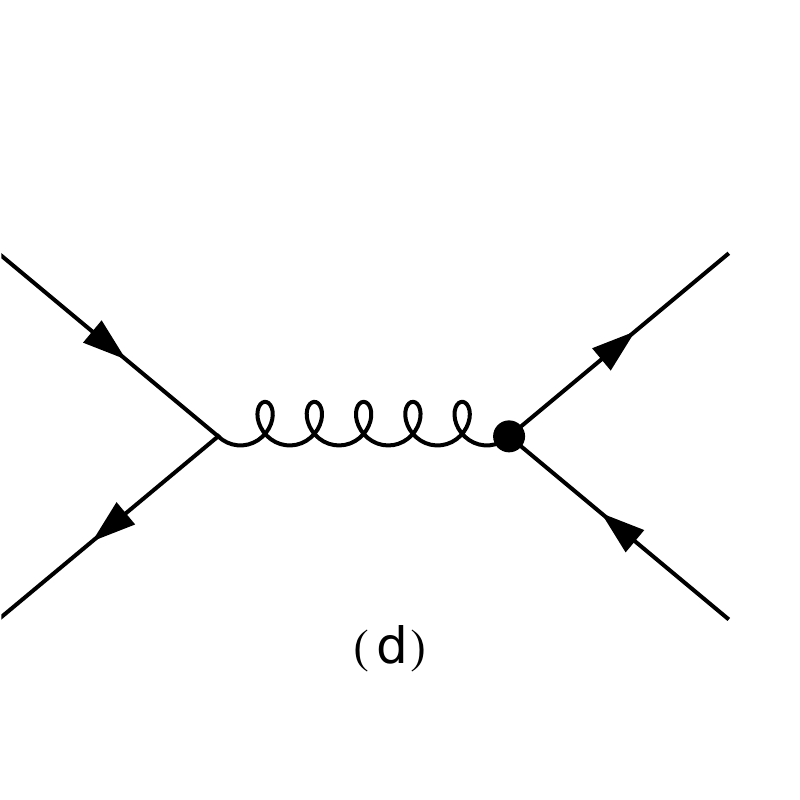}
\caption{\small Representative Feynman diagrams for $t\bar t$ production through gluon fusion (a)-(c) and quark-antiquark annihilation (d).  The black dot represent the insertion of one of the two operators of Eq.~\eqref{q1} and Eq.~\eqref{q2}.}
\label{fig:ttbar-diagrams} 
\end{center}
\end{figure}

We update here these constraints by means of the most precise 13 TeV LHC data.
The CMS collaboration recently released a measurement of the top quark pair total cross section performed in the single lepton channel with an integrated luminosity of $3.2\;$fb$^{-1}$~\cite{Sirunyan:2017uhy}. This measurement yields a value for the total cross section of 
\begin{equation}
\sigma(pp\to t \bar t)=835\pm 3\;(\rm stat) \pm 23\;(\rm syst) \pm 23\;(\rm lum)\;{\rm pb}.
\end{equation}

The relative error on this measurement, after having summed in quadrature the various sources of uncertainty, is about 3.9\%, comparable to the one obtained with the combination of 7 and 8 TeV data in the dileptonic channel~\cite{Khachatryan:2016mqs}.

\begin{figure}[ht!]
\begin{center}
\includegraphics[width=0.45\textwidth]{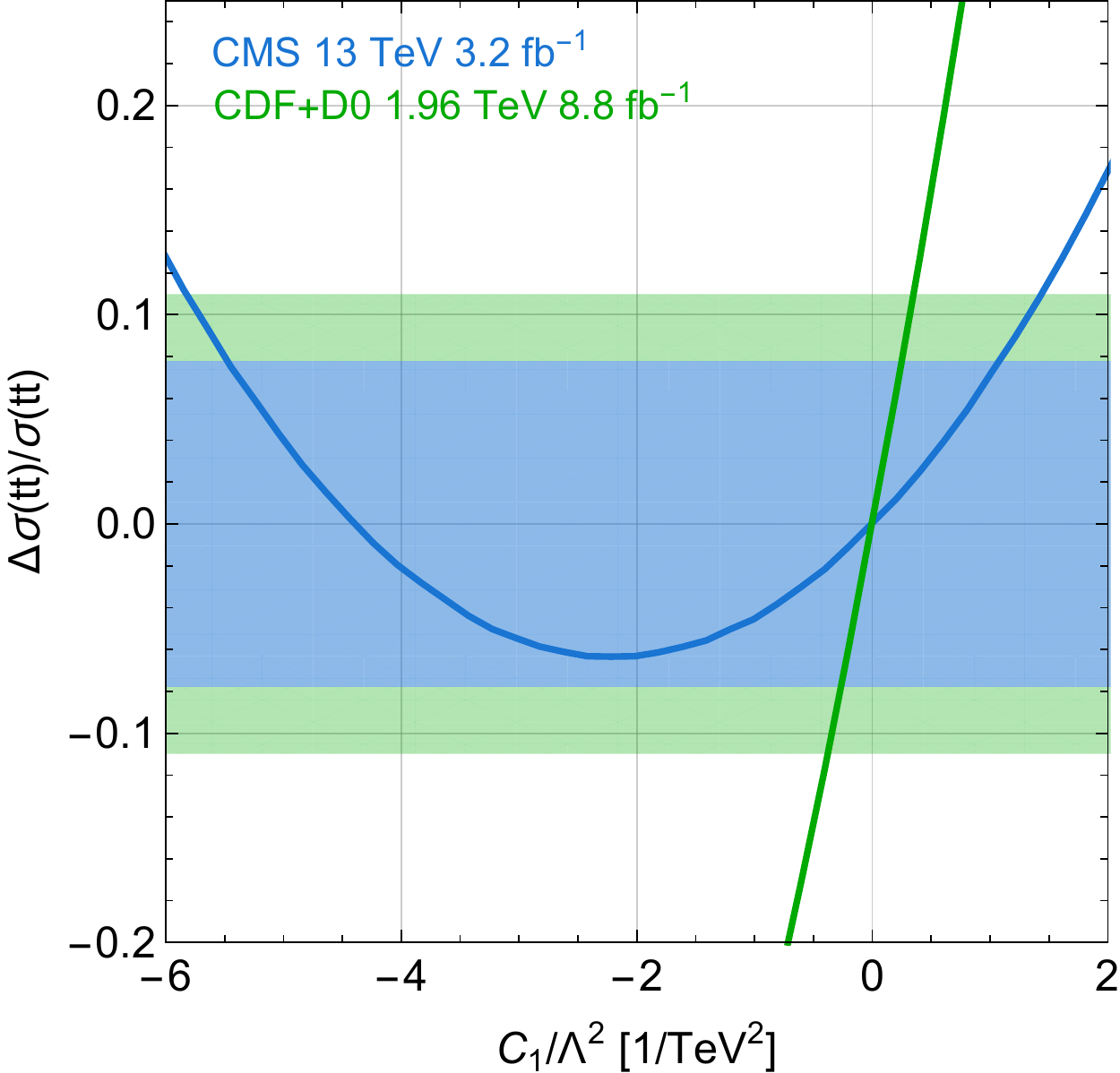}
\vskip0.4cm
\includegraphics[width=0.45\textwidth]{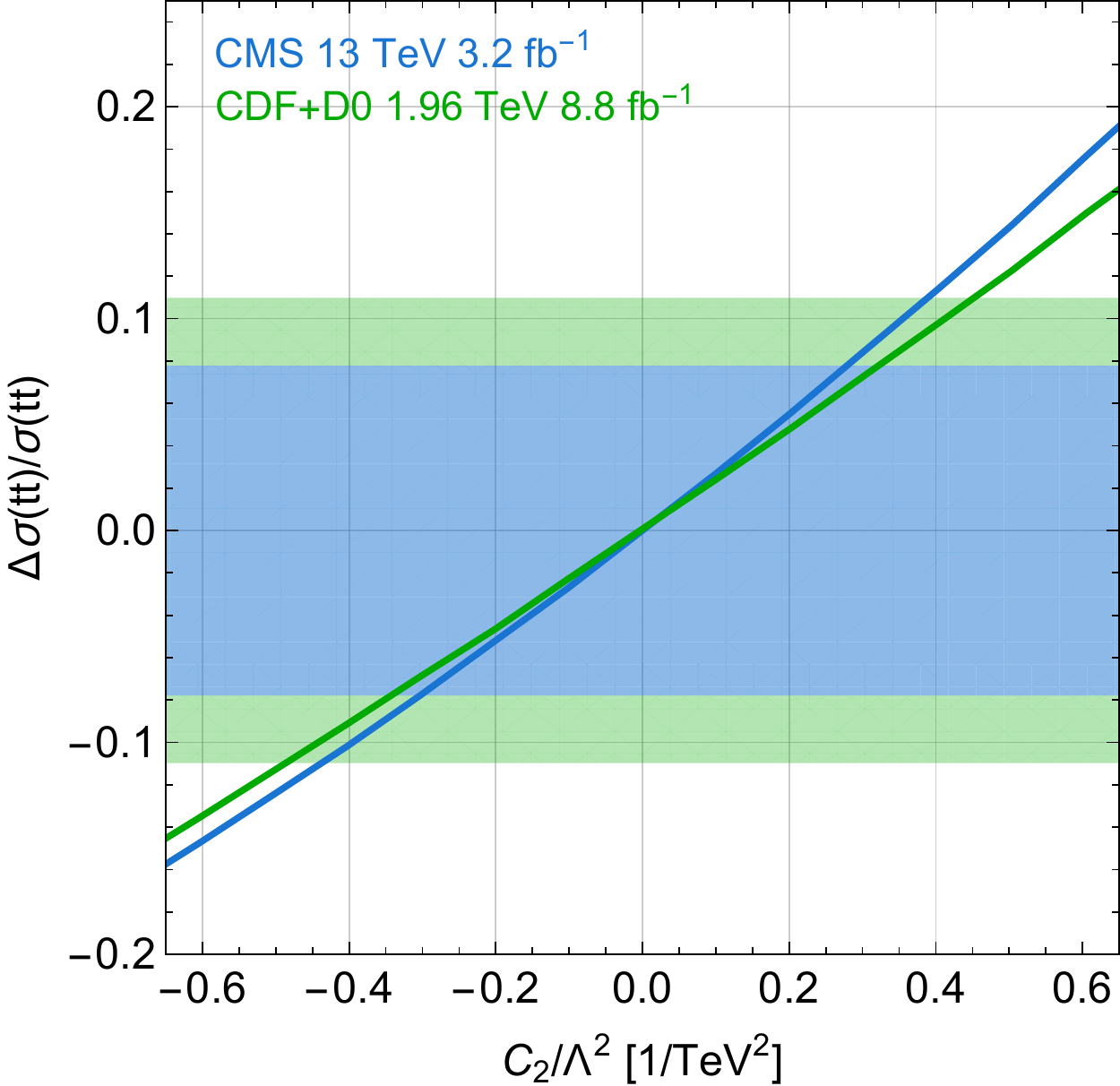}
\caption{Relative modification of the $t\bar t$ total cross section, $\Delta\sigma(t\bar t)/\sigma(t\bar t)=\sigma(t\bar t)^{\rm BSM}/\sigma(t\bar t)^{\rm SM}-1$, induced by the presence of the operator $\mathcal O_1$ and $\mathcal O_2$. The blue and green shaded regions correspond to the 95\% confidence level intervals on the Wilson coefficient $C_1$ and $C_2$ from the cross section determination from LHC and Tevatron data respectively.  The limits can be found by looking at  the intersections of the  curves with the regions of the same color: 
$-5.48/{\rm TeV^2}<C_1<1.08/{\rm TeV^2}$ and $-0.30/{\rm TeV^2}<C_2<0.28/{\rm TeV^2}$ for the LHC and $-0.38/{\rm TeV^2}<C_1<0.35/{\rm TeV^2}$ and $-0.49/{\rm TeV^2}<C_2<0.45/{\rm TeV^2}$ for Tevatron.}
\label{fig:exclusive}
\end{center}
\end{figure}

The operator of Eq.~\eqref{q1} does not affect the partonic process $gg\to t\bar t$, thus only modifying  the $q\bar q$ initiated reaction, which at the LHC is subdominant in the $t\bar t$ cross section, given that the anti-quark parton has to be extracted from the sea quarks of the proton. This comes about because of gauge invariance and the presence of a contact vertex with two gluons attached to the quark lines (see Fig.~\ref{fig:ttbar-diagrams} (a) and (c)), a contribution which cancels out that of the vertex with a single gluon.
For this reason the Wilson coefficient $C_1$ can be more effectively constrained by Tevatron data, where the anti-quark state is  extracted from the valence quarks of the colliding anti-proton.  
A combined results from the CDF and D0 collaboration gives the following measurement of the total $t\bar t$ cross section~\cite{Schilling:2013nca}
\begin{equation}
\sigma(p\bar p\to t \bar t)=7.65 \pm 0.42\;{\rm pb}
\end{equation}
with a relative precision of about 5.5\%, which we use throughout our analysis.

\begin{figure}[ht!]
\begin{center}
\includegraphics[width=0.45\textwidth]{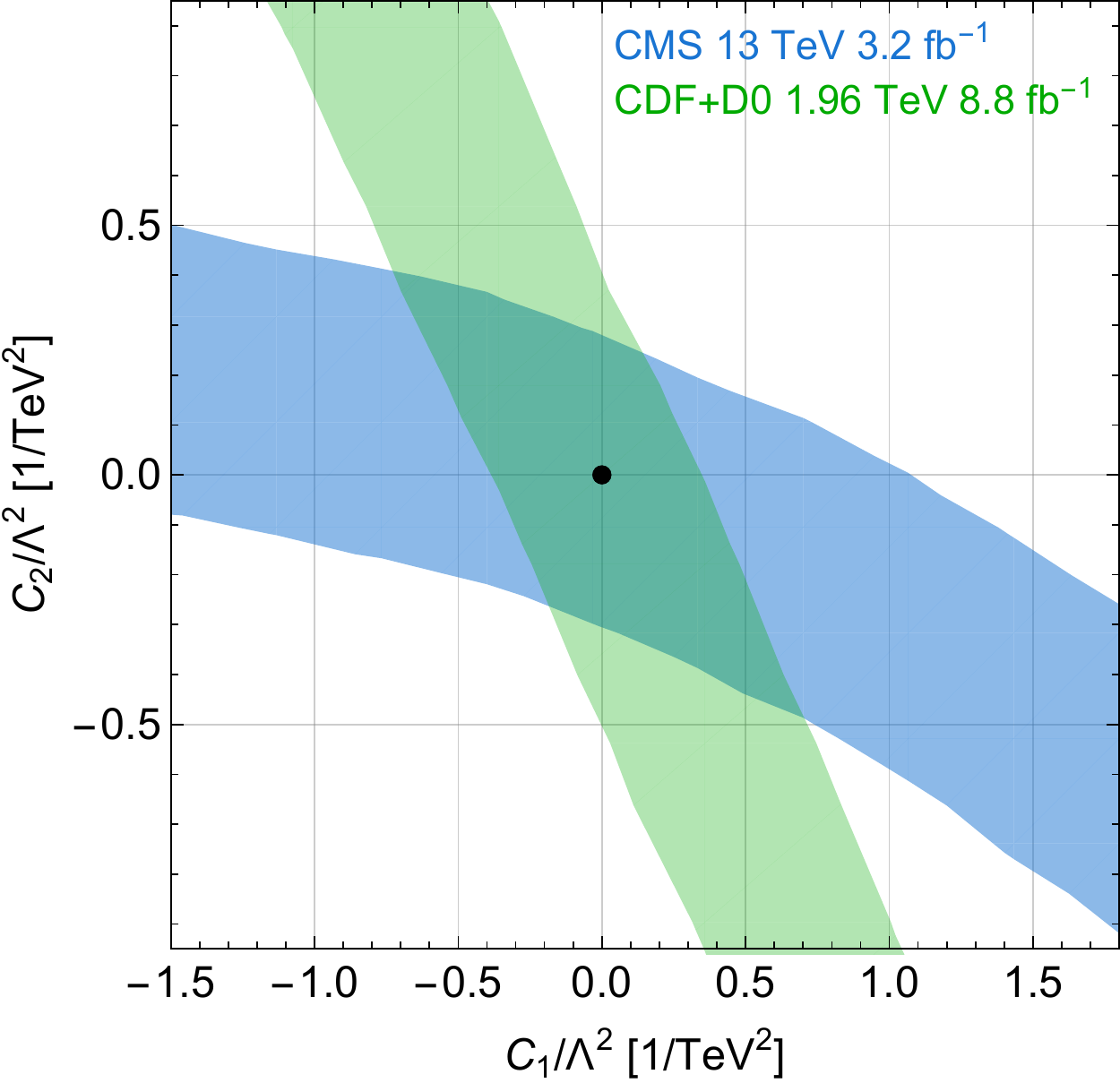}\hfill
\caption{95\% confidence intervals on the Wilson coefficient  $C_1$ and $C_2$ from the measurements of the $t\bar t$ total cross section at the LHC, blue, and Tevatron, green. The corresponding combined limits are listed in Table~\ref{tab:current}.}
\label{fig:combined}
\end{center}
\end{figure}

Following the procedure described at the beginning of this Section, and by fixing one of the two Wilson coefficient to zero, we obtain the limits from the total $t\bar t$ cross sections measurements which are shown in Fig.~\ref{fig:exclusive}, where the blue and green shaded areas correspond to the 95\% confidence level uncertainties on the cross section determination at the LHC and Tevatron respectively, and the solid lines correspond to the relative modification of the SM cross section due to the presence of the operator $\mathcal O_1$ and $\mathcal O_2$.  If we allow for the presence of both operators at the same time, we obtain the limits shown in Fig.~\ref{fig:combined}. The two exclusion regions have different inclination because the operator $\mathcal O_1$ contribution depends on the different relative importance of the gluon and quark initiated reaction at the Tevatron and the LHC. The importance of the Tevatron data in constraining the $C_1$ Wilson coefficient is thus manifest, the bound being a factor three better for $C_1>0$ (taking $C_2=0$).

\subsection{Limits from the differential  cross sections}
\label{sec:tt}

\begin{figure}[ht]
\begin{center}
\includegraphics[width=0.45\textwidth]{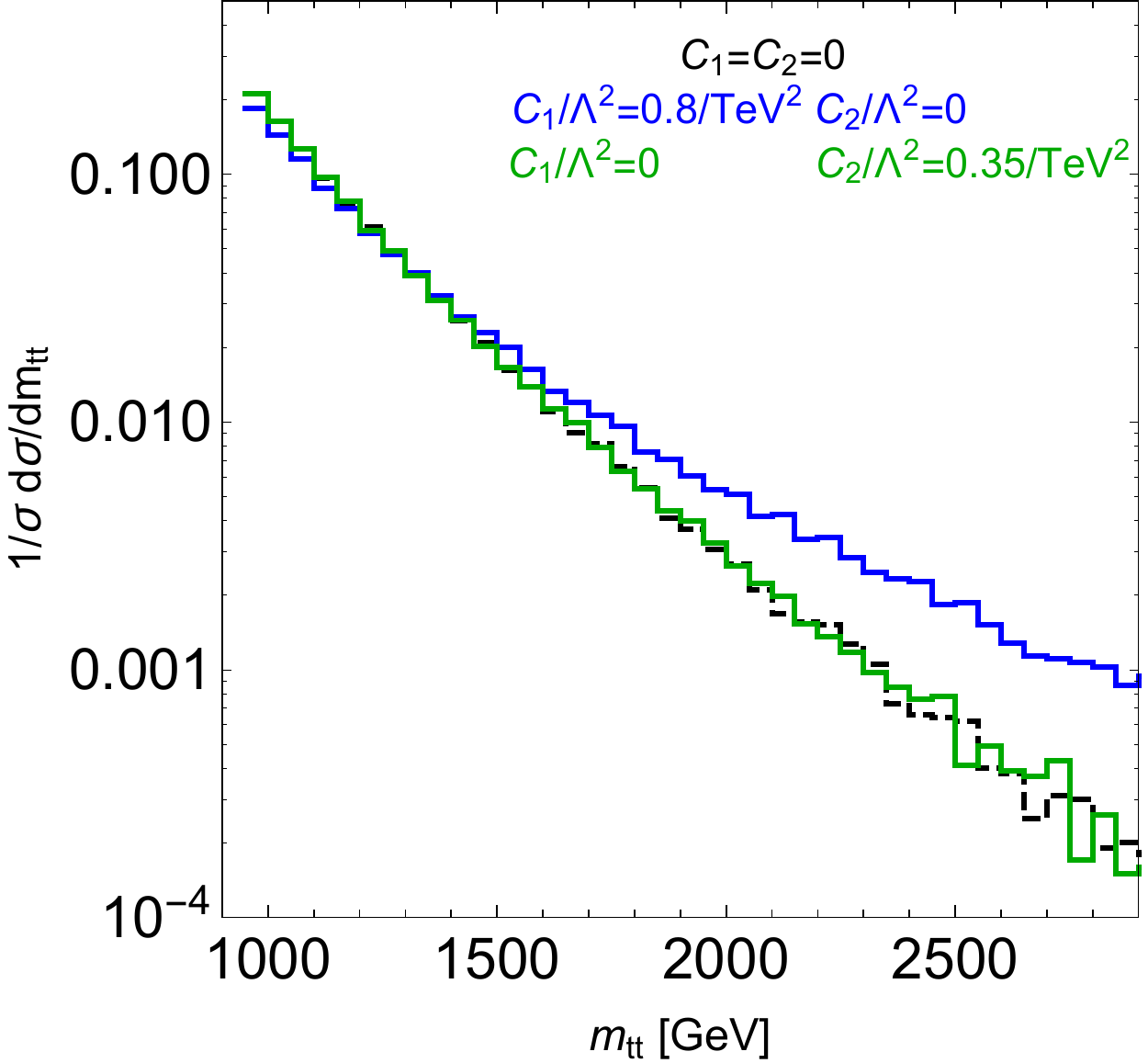}
\vskip 0.4cm
\includegraphics[width=0.45\textwidth]{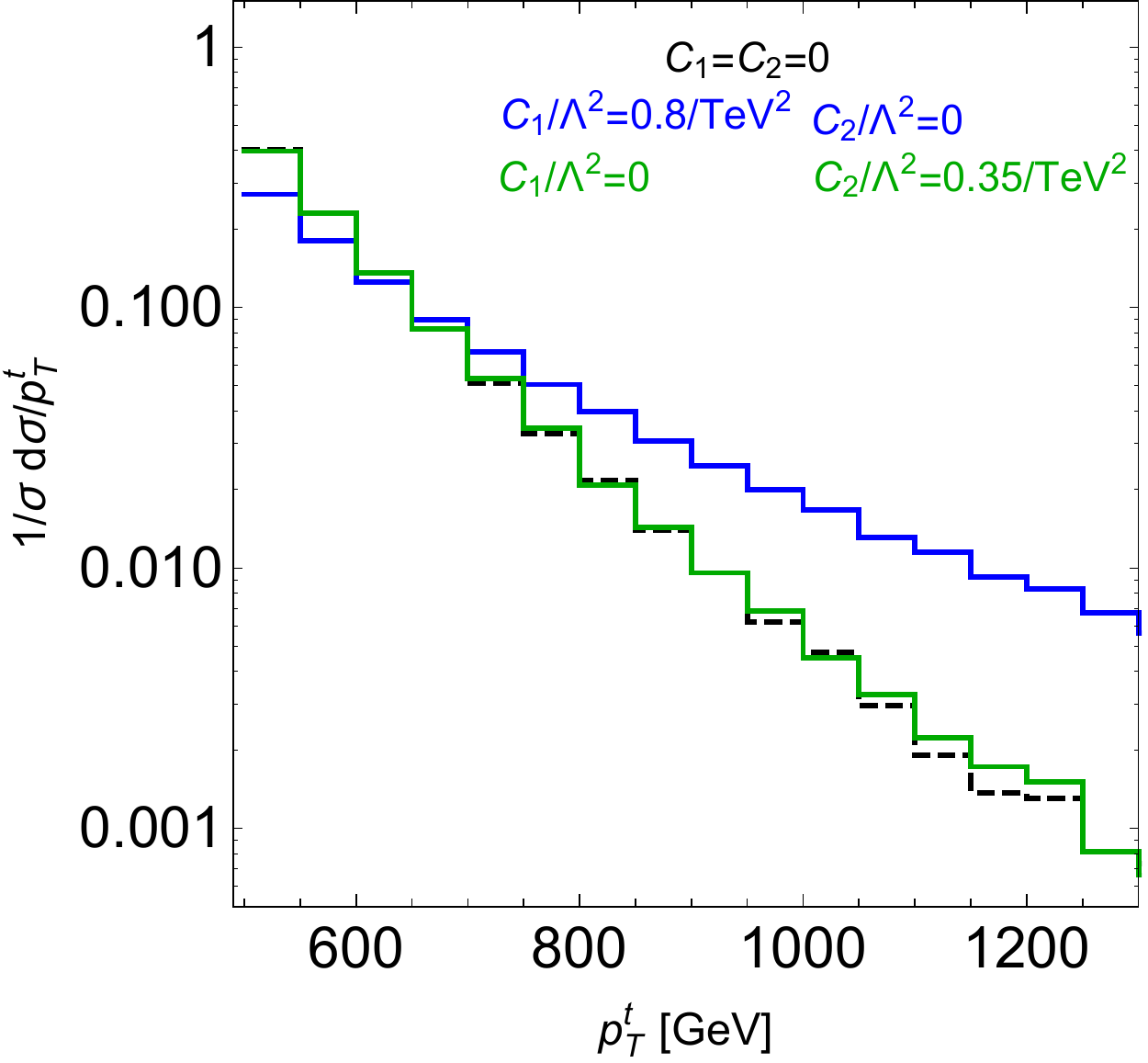}
\caption{ Differential distribution for top quark pair production with respect to the top pair invariant mass and the top transverse momentum normalized to unity for the case where just the operator ${\cal O}_1$ is inserted, solid blue, and just the operator ${\cal O}_2$ is inserted, solid green. The SM prediction is shown in dashed black. The relative independence from ${\cal O}_2$ is manifest. Also, it is for large $m_{t\bar t}$ and $p^t_T$ that the distributions are most sensitive  to the insertion of  ${\cal O}_1$.}
\label{fig:shape}
\end{center}
\end{figure}

The current center of mass energy for LHC proton collisions and the large number of top quark pairs expected to be produced during the present run of the CERN machine, will allow to measure top quark differential cross sections with an unprecedented precision and with a potential large number of events populating the tails of such distributions, thus allowing for a more stringent comparison between experimental measurements and theoretical predictions. 
In fact, other than modifying the total rate for $t\bar t$ production, the effective operators  in Eq.~\eqref{q1} and Eq.~\eqref{q2} can in principle affect the shape of the cross sections differential distributions, altering them with respect to the SM predictions. 
Therefore the possibility of using {\emph{differential measurements}} other than total cross sections potentially offers a powerful mean to constrain the coefficients of these higher dimensional operators.

In particular, both the top quark pair invariant mass differential distribution ($\di \sigma/\di m_{t\bar t}$) and the top quark transverse momentum differential distribution ($\di \sigma/\di p^t_{T}$) present an interesting behavior with respect to the two operators of Eq.~\eqref{q1} and Eq.~\eqref{q2}.  The insertion of the ${\cal O}_1$ operator gives rise to the typical tail enhancement in the distributions at large invariant masses and transverse momentum, as shown in Fig.~\ref{fig:shape}, where the differential rates normalized to the total cross section are computed for both the case of the top pair invariant mass distribution and top quark transverse momentum. 

On the other hand, for high invariant masses and transverse momenta, the shapes of the differential distributions computed in presence of the operator ${\cal O}_2$ are not modified with respect to the SM when just the linear order in the Wilson coefficient $C_2$ is retained. This is true at LO~\cite{Zhang:2010dr} but also at NLO, as shown in~\cite{Franzosi:2015osa}, where both the SM and the EFT contributions are evaluated at NLO order. The computation of~\cite{Franzosi:2015osa} shows that evaluating both terms at NLO order avoids an overestimation of the enhancement of the contribution of the ${\cal O}_2$ operator in the high energy regime.

The inclusion of quadratic terms in $C_2$ modifies the high energy tails of the distributions already at tree level. In~\cite{Aguilar-Saavedra:2014iga}, the authors retain up to quartic terms in the effective operator coefficients for computing the cross sections and find an enhancement of the sensitivity in the ultra boosted regime. However,
the contribution of these quadratic terms is negligible if the specific values of $C_2$ used to generate the distributions  in the relevant energy range are sufficiently small (see  Fig.~\ref{fig:shape}).

For what concerns our analysis, the different behavior of the two operators suggests that the \emph{normalized} differential cross section measurements can be used to set a limit on the coefficient of the ${\cal O}_1$ operator, irrespective of the value taken by the ${\cal O}_2$ operator.

From the experimental side, while the invariant mass distribution of the top quark pairs, $m_{t\bar t}$, has been previously measured by both the CDF and D0 collaboration at Tevatron~\cite{Aaltonen:2009iz,Abazov:2014vga}, more recently both the ATLAS and CMS collaborations have provided unfolded measurements of this and others observable both normalized to the total event rate and to unity~\cite{ATLAS:2016soq,CMS-PAS-TOP-16-007,CMS-PAS-TOP-16-013,ATLAS:2016jct,Aaboud:2016syx}. We will use the ATLAS differential measurements of~\cite{ATLAS:2016jct} which have been performed in the all hadronic channel with an integrated luminosity of 14.7 fb$^{-1}$ exploiting a final state with highly boosted top, which have been shown to be effective in testing the top quark intrinsic structure~\cite{Aguilar-Saavedra:2014iga}. 

\begin{table}[ht]
\begin{tabular}{c || c | c | c | c | c | c | c | c | c }
$m_{t\bar t}$ [TeV] & 1.0 & 1.1 & 1.2 & 1.3 & 1.4 & 1.5 & 1.7 & 2.0 & 2.3-3.0  \\
\hline
Error [\%] 	            & 36 & 20 & 25 & 30 & 31 & 32 & 63 & 58 & 123   \\
\end{tabular}
\vskip 10pt
\begin{tabular}{c || c | c | c | c | c | c  }
$p_{T}^t$ [TeV] & 0.5 & 0.55 & 0.6 & 0.65 & 0.75 & 0.9-1.2   \\
\hline
Error [\%] 	            & 19 & 25 & 28 & 45 & 73 & 95     \\
\end{tabular}
\caption{Top pair invariant mass and top quark transverse momentum binning of the ATLAS measurements of $t\bar t$ invariant mass differential cross section and relative errors in \%~\cite{ATLAS:2016jct}. The observable values indicate the lower edge of the considered bin except for the last bin where the upper values value are explicitly indicated.}
\label{tab:binning}
\end{table}

\begin{figure}[ht!]
\begin{center}
\includegraphics[width=0.45\textwidth]{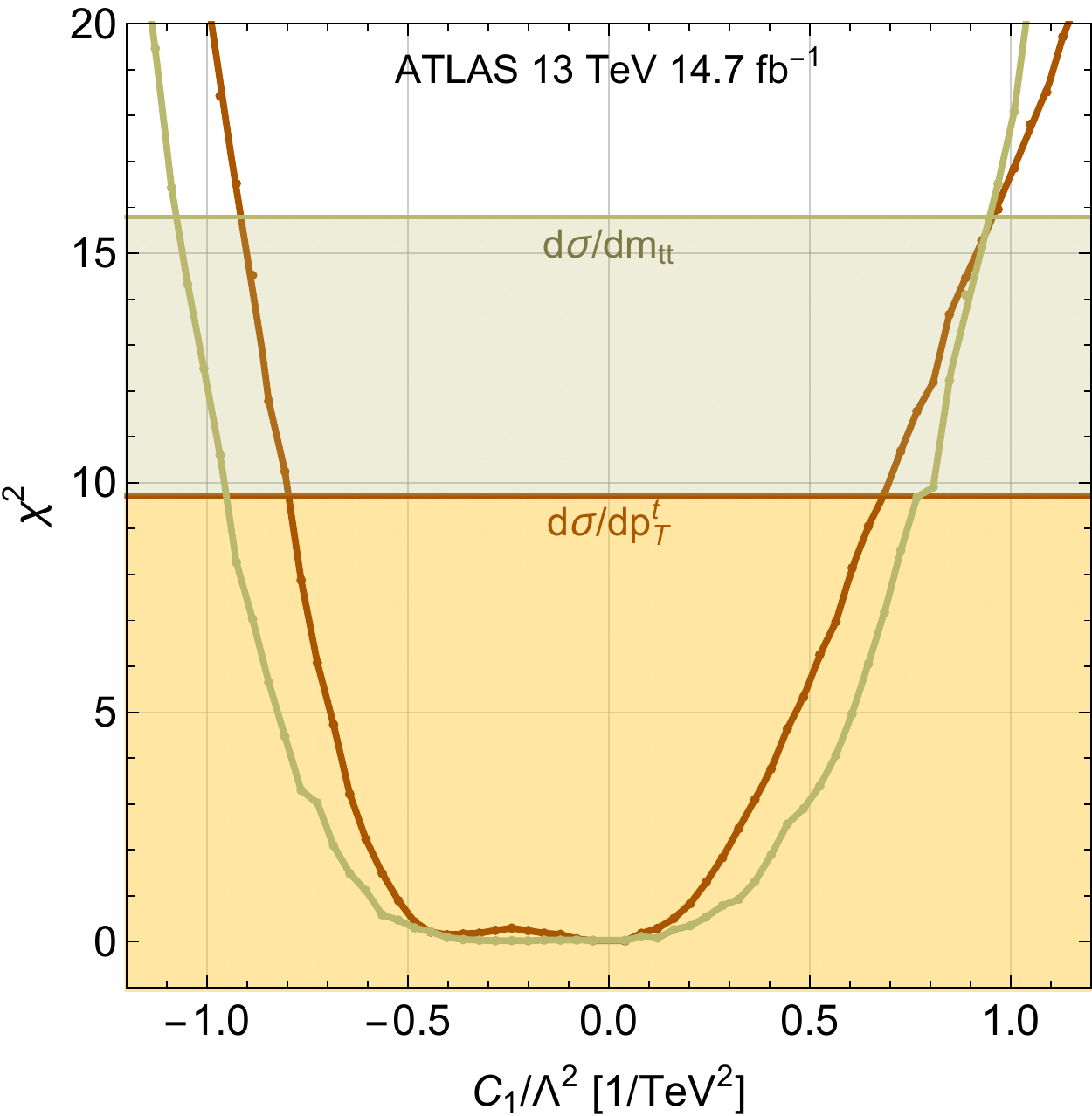}
\caption{$\chi^2$ distribution for the Wilson coefficient $C_1$ from the differential cross section measurement in $t\bar t$ invariant mass and top quark transverse momentum of~\cite{Aaboud:2016syx}. The horizontal lines represent the 95\% confidence level limit taken for a $\chi^2$ with 8 degrees of freedom corresponding to the 9 bins of data considered in the $1/\sigma\;\di\sigma/\di m_{t\bar t}$ distribution and  5 degrees of freedom corresponding to the 6 bins of data considered in the $1/\sigma\;\di\sigma/\di p^t_{T}$ distribution.}
\label{fig:chi}
\end{center}
\end{figure}

We thus perform a $\chi^2$ fit to the measured top quark normalized invariant mass and transverse momentum distributions, see Fig.~\ref{fig:chi}, again assuming that the central value of the experimental measurements coincides with our predictions when $C_1=C_2=0$, with the uncertainties reported in Tab.~\ref{tab:binning}. The number of degrees of freedom for the $\chi^2$ fit correspond to the number of bins of the considered distribution minus one, since one degree of freedom is fixed by the requirement that the area under the curve is equal to unity. With this procedure, and taking the data for the $p^t_T$ distribution which turn out to provide the most stringent constraint, we set  a limit of $-0.80/{\rm TeV^2}<C_1/\Lambda^2<0.68/{\rm TeV^2}$, which is comparable with the one that can be obtained through the total cross section measurements. We will  show in Sec.~\ref{sec:combine} the prospects for the determination of the $C_1$ coefficient with the increase of the data collected by the LHC.

\section{Higgs production cross section measurements}
\label{sec:higgs}

\begin{figure*}[ht!]
\begin{center}
\includegraphics[width=0.22\textwidth]{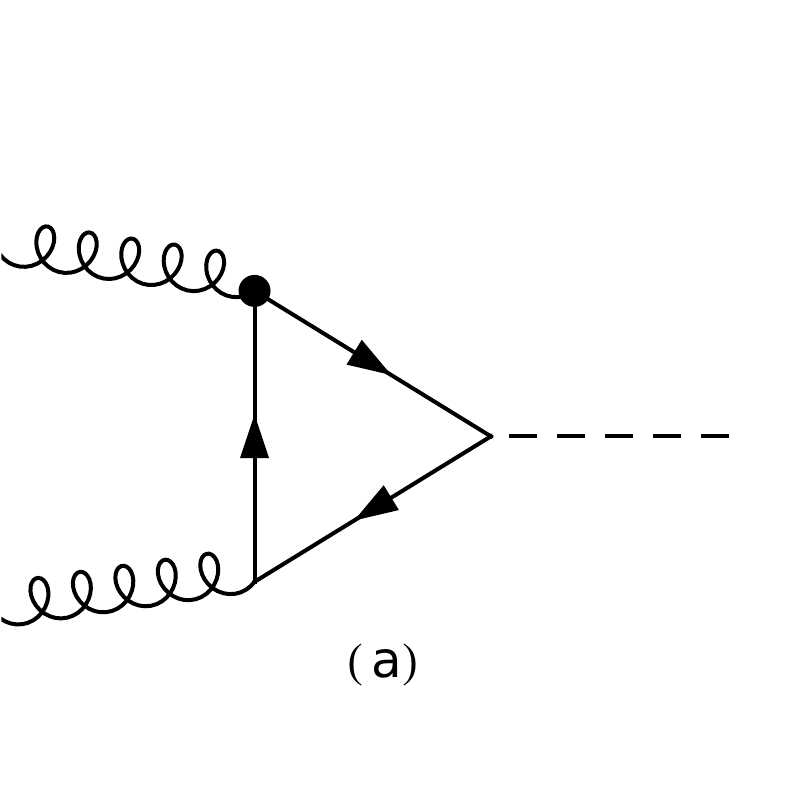}
\includegraphics[width=0.22\textwidth]{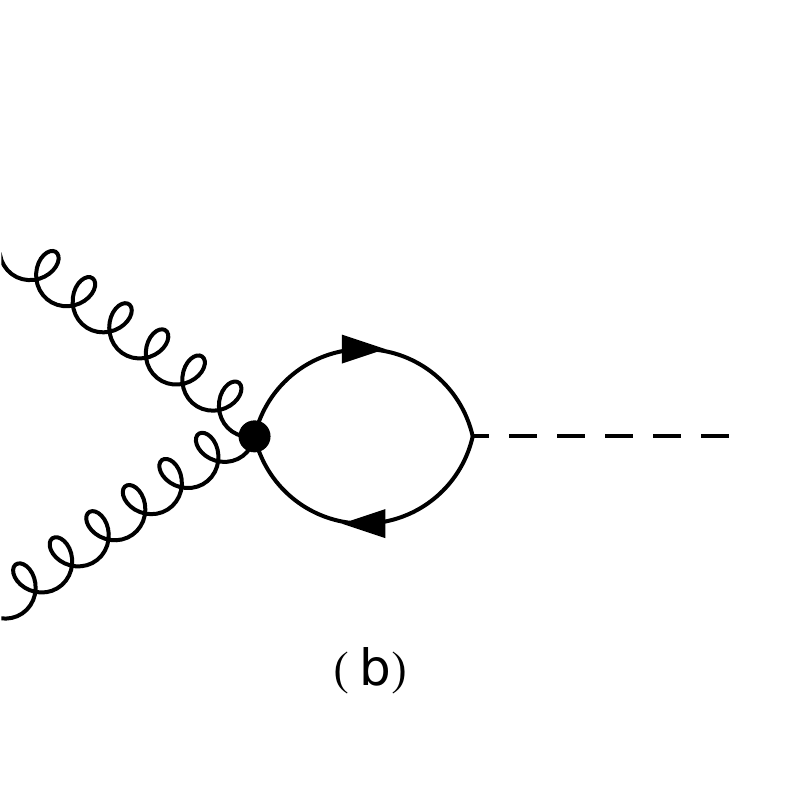}
\includegraphics[width=0.22\textwidth]{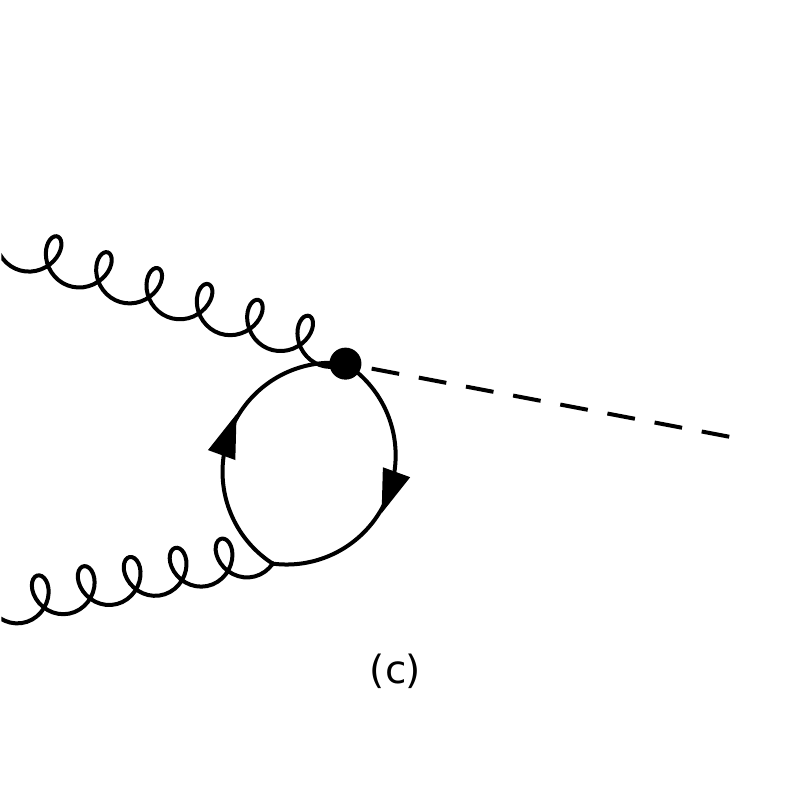}
\caption{\small Representative Feynman diagrams for Higgs boson production through gluon fusion. The black dot represents the insertion of the operator of Eq.~\eqref{q2}.}
\label{gghdiagrams}
\end{center}
\end{figure*}

The production of the Higgs boson at the LHC is dominated by the gluon fusion channel. This process arises in the SM from a one loop diagram mediated by colored fermions, the amplitude being dominated by the top quark contribution because of its large Yukawa coupling to the Higgs boson. It has been discussed as an observable sensitive to the operator ${\cal O}_2$ in \cite{Degrande:2010kt,Chien:2015xha}.

The presence of the higher dimensional operators of Eq.~\eqref{q1} and Eq.~\eqref{q2} introduces modifications to the coupling between the top quark and the gluon thus affecting the Higgs boson production rate.
Only the ${\cal O}_2$ operator  contributes to this process because for on-shell gluons the correction arising from ${\cal O}_1$ identically vanishes. This is obvious if one recalls that the operator ${\cal O}_1$ can be written in terms of four-fermion operators, as shown in Appendix~\ref{appendix}.
 Therefore the amplitude for $gg\to H$ can be written as the sum of two contributions
\be
{\cal M}={\cal M}_{\rm SM}+{\cal M}_{{\cal O}_2}\,,
\label{eq:higgs-amplitude}
\ee
where ${\cal M}_{\rm SM}$ is the SM contribution and ${\cal M}_{\rm {\cal O}_2}$ is the contribution coming from one insertion of the ${\cal O}_2$ operator.
Terms coming from two insertion of the dipole operator are neglected, since in the end we are going to retain only contributions linear in $C_2$, as discussed in section~\ref{sec:fine_print}.
We assume that the Yukawa coupling between the top quark and the Higgs boson takes its SM value and we take a zero finite contribution from the operator ${\cal O}_{HG}=H^\dag H G^a_{\mu\nu} G^{\mu\nu}_a$. Furthermore we assume that their mixing with the operators of Eq.~\eqref{q1} and Eq.~\eqref{q2} is negligible. With these assumptions we can use the $gg\to H$ process to set a direct limit on the coefficient of the ${\cal O}_2$ operator.
We rewrite the effective operator of Eq.~\ref{q2} in its $SU(2)_L \times U(1)_Y$ invariant form in order to correctly take into account all the contributions affecting Higgs phenomenology arising from the operator $\mathcal O_2$, see Fig.~\ref{gghdiagrams}.

We compute the Higgs production cross section analytically, cross checking the results by means of {\tt Package X}~\cite{Patel:2015tea}. The final numerical integration of the Feynman integrals has also been checked against {\tt FormCalc8}~\cite{ChokoufeNejad:2013qja}. 
A factor 4 takes into account the identical contributions coming from crossing of the gluon lines and switching the vertex insertion of the dipole operator. The contribution of the diagram (b) of Fig.~\ref{gghdiagrams} turns out to be identically zero in dimensional regularization. 
 We therefore have
\begin{widetext}
\begin{equation}
\begin{split}
& ({\cal M}_{\mathcal O_2})^{ab}_{\lambda_1 \lambda_2}  =  4 \times g_s \frac{m_t}{\sqrt{2}}\frac{2\;C_2}{\Lambda^2}  \frac{1}{16\pi^2}(m_H^2 g_{\mu\nu} - 2 q_{2\mu} q_{1\nu}) \varepsilon^\mu_{\lambda_1} (q_1) \varepsilon^\nu_{\lambda_2} (q_2) \, \Tr \left[T^a T^b \right]  
\times  \left\{ \frac{1}{\bar \epsilon} + 1-
\log \frac{\mu^2}{m_t^2}  \right.  \\
& \left.  + \frac{m_t^2}{m_H^2}  \log^2 \left( \frac{ \sqrt{m_H^4 - 4 m_t^2 m_H^2} + 2 m_t^2 - m_H^2}{2 m_t^2} \right) +
\frac{\sqrt{m_H^4 - 4 m_t^2 m_H^2}}{m_H^2} \log \left( \frac{\sqrt{m_H^4 - 4 m_t^2 m_H^2} + 2 m_t^2 - m_H^2}{2 m_t^2}\right) \right\} \, 
\end{split}
\end{equation}
\end{widetext}
where $m_t$ and $m_H$ are the masses of the top quark and the Higgs boson. The vectors $\varepsilon^\mu_{\lambda_1}(q_1)$ and $\varepsilon^\nu_{\lambda_2} (q_2)$ represent the polarizations for the two incoming gluons with momenta $q_1$ and $q_2$. We regularize the divergent loop integral by means of dimensional regularization  where the pole in 4 dimensions is written in the $\overline{MS}$ scheme, {\it i.e.} $ 1/\bar \epsilon= 1/\epsilon-\gamma_E+\log(4\pi)$.

 In order to have a finite amplitude we subtract the $ 1/\bar \epsilon$ pole by a counter-term proportional to the effective operator describing the direct coupling of the Higgs boson to the gluon fields: ${\cal O}_{HG}=H^\dag H G^a_{\mu\nu} G^{\mu\nu}_a$. This renormalization procedure leaves a logarithmic dependence on the subtraction  scale, which we take  $\mu=m_H$ to match the factorization scale for the process. 
 We also explicitly checked that indeed the double insertion of the $\mathcal O_2$ operator gives rise to a  small correction that can be neglected, as discussed in section~\ref{sec:fine_print}.

\begin{figure}[ht!]
\begin{center}
\includegraphics[width=0.45\textwidth]{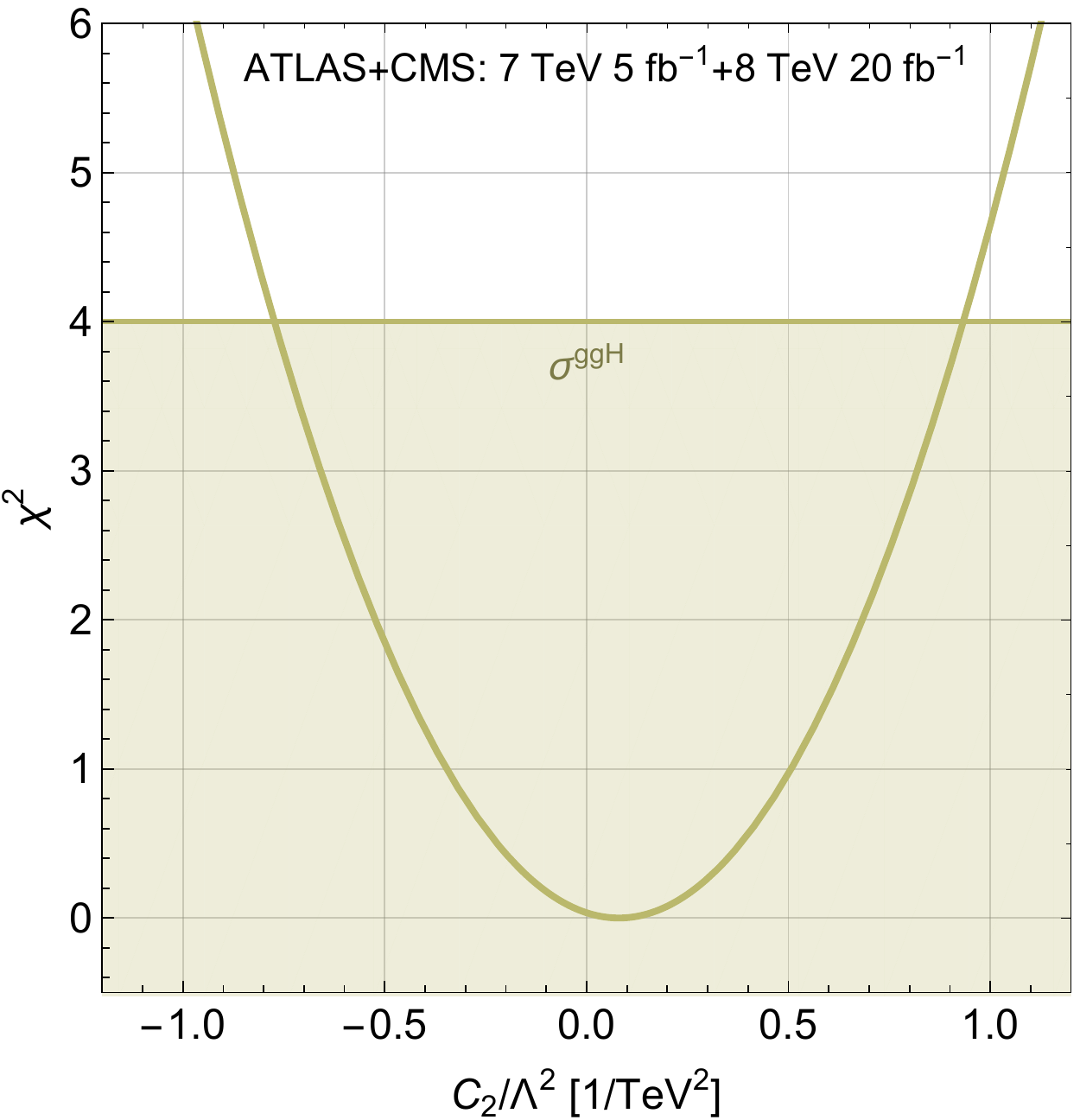}
\caption{
$\chi^2$ distribution for the Wilson coefficient $C_2$ from the Higgs production from gluon fusion.
The horizontal line represents the 95\% confidence level limit taken for a $\chi^2$ with 1 degrees of freedom.
\label{fig:chi2} }
\end{center}
\end{figure}

In computing the squared amplitude of Eq.~\eqref{eq:higgs-amplitude} the leading correction to the SM cross section is a term linear in the $C_2$ Wilson coefficient. By fixing $m_{t}=172\;$ GeV and $m_H=125\;$ GeV we find that the ratio of the gluon fusion Higgs production cross section with respect to its SM value is
\be
\mu_{{\cal O}_2} \simeq 1 + 0.375~{\rm TeV^2} \frac{C_2}{\Lambda^2} \,.
\label{eq:higgs-mod}
\ee
This ratio is measured experimentally and usually presented by the experimental collaborations either in terms of {\emph signal strengths} values, which is precisely the ratio of the experimental measurements with respect to the SM expectation, or  of coupling modifier, the ratio of the Higgs to gluon gluon effective coupling compared with the SM prediction. In either cases, the results of Eq.~\eqref{eq:higgs-mod} allows us to directly use the current precision on the Higgs production measurements and  set a limit on the $C_2$ Wilson coefficient.

As for the computation of the $t\bar t$ production cross section, this ratio has been obtained at LO. We however assume this results to hold also at NLO  since the $k$ factor induced by higher order corrections are expected to be the same for the SM and the effective operator cases, therefore cancelling out in performing the ratio.

The ATLAS and CMS collaborations have performed a combined measurements of the Higgs signal strength with  about 5  and  20 fb$^{-1}$ of data collected during the 7 and 8 TeV run of the LHC, yielding a value for the gluon fusion Higgs production signal strength~\cite{TheATLASandCMSCollaborations:2015bln}
\be
\mu_{ggH} = 1.03^{+0.17}_{-0.15}.
\ee

The $\chi^2$ value for the parameter $C_2$ is shown in Fig.~\ref{fig:chi2} from which we find the 95\% confidence level limits $-0.77/{\rm TeV^2}<C_2/\Lambda^2<0.93/{\rm TeV^2}$, also reported in Table~\ref{tab:current}. This estimate provides limits on the coefficient of the ${\cal O}_2$ operator,  which are not yet competitive with those obtained from the measurements of the top pair production cross section.
We will show in the next Section how the expected improvement on the determination of this signal strength will provide  stronger limits on the $C_2$ Wilson coefficient.

\section{Combination and prospects}
\label{sec:combine}

\begin{figure}[h!]
\begin{center}
\includegraphics[width=0.45\textwidth]{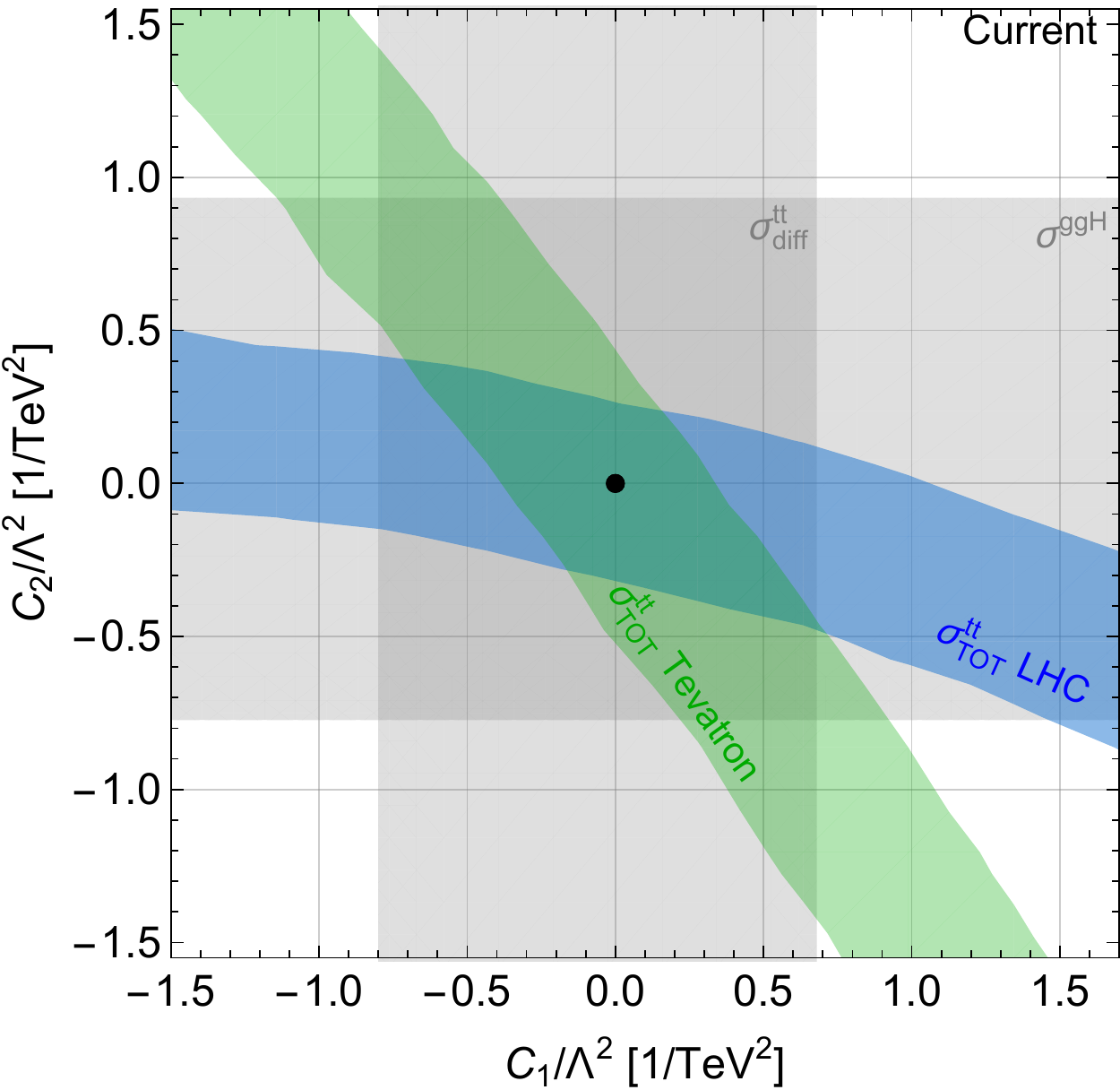}\hfill
\caption{95\% confidence intervals for  $C_1$ and $C_2$ from the measurements of the top quark transverse momentum differential cross section and Higgs production via gluon fusion cross section, vertical gray and horizontal gray shaded area respectively, with current available data. The limits from the measurements of  $t\bar t$ total cross section at the LHC (blue) and Tevatron (green) are also shown.}
\label{fig:results_present}
\end{center}
\end{figure}

In the previous sections we have shown that the measurement of the normalized top quark transverse momentum differential distribution in top pair production and the measurement of the Higgs boson production cross section through gluon fusion can be used to set \textit{independent limits} on the coefficient of the operators ${\cal O}_1$ and ${\cal O}_2$ respectively. We  show in Fig.~\ref{fig:results_present} the limits on the $C_1$ and $C_2$ Wilson coefficient obtained through this method, together with those obtained only by means of the measurements of the total top pair pair production cross sections performed at both Tevatron and LHC. Table~\ref{tab:current} summarizes the various bounds. 
These bounds are the most stringent among those so far available for the operators ${\cal O}_1$ and ${\cal O}_2$ (compare with those in \cite{Fabbrichesi:2013bca} and \cite{Buckley:2015nca}).

\begin{table*}[ht!]
\begin{center}
\vspace{0.2cm}
\begin{tabular}{|c|c|c|}
\hline
 $\sigma_{t\bar t}$ (Tevatron + LHC)  &  $\mu_{ggH}$ & $ \di \sigma_{t \bar t}/\di p_{T}^t$  \cr  
\hline
\hline
$\qquad -0.74 < C_1/\Lambda^2 < 0.71 \qquad$    & --- & $\qquad -0.80 < C_1/\Lambda^2 < 0.68 \qquad $ \cr
\hline
$\qquad -0.49 < C_2/\Lambda^2 < 0.42 \qquad$   & $\qquad -0.77 < C_2/\Lambda^2 < 0.93 \qquad $ & --- \cr
\hline
\end{tabular}
\end{center}
\caption{Limits at 95\% confidence level on the coefficients  $C_{1}$ and $C_2$ from current data. Values in the first column come from the total cross sections and are obtained by marginalization of one operator against the other. The limits in the next two columns are obtained for the two operators independently by means of Higgs production and the indicated differential cross section.  All values are in units of TeV$^{-2}$.}
\label{tab:current}
\end{table*}

The proposed method thus sets limits comparable to those obtained from total $t\bar t$ cross section measurements on the operator $\mathcal O_1$ and roughly a factor two weaker on the operator $O_2$.
However, while the current uncertainties on the measurement of the top quark pair total cross section, which  are about 4\%, are not going to  improve substantially, this is not the case for the top quark differential cross sections as well as for  the Higgs production cross section measurements which are expected to become more precise.
In order to infer the projected limits on the $C_1$ and $C_2$ Wilson coefficients we thus proceed in the following way.

For the measurements of the top quark transverse momentum differential cross section we rescale the uncertainties reported in Tab.~\ref{tab:binning} by the luminosity dependent factor $\sqrt{\mathcal L_0/\mathcal L}$ where
$\mathcal L_0=14.7$ fb$^{-1}$ indicates the current collected luminosity and $\mathcal L$ the projected luminosity. We finally take the error associated with this measurement to be
\begin{equation}
\frac{\Delta \sigma}{\sigma}\bigg\rvert_\mathcal L = {\rm Max}\left[ 0.15,\frac{\Delta \sigma}{\sigma}\bigg\rvert_{\mathcal{L}_0}\times \sqrt{\frac{\mathcal L_0}{\mathcal L}} \right]
\end{equation}
thus assuming a conservative floor of 15\% for the error estimation.

For the Higgs production through gluon fusion process, we use the projected uncertainties on the measurements as provided by the CMS collaboration~\cite{CMS-NOTE-2012-006} which are 5.7\% (2.7\%) for a collected integrated luminosity of 300 (3000) fb$^{-1}$.

\begin{figure}[ht!]
\begin{center}
\includegraphics[width=0.45\textwidth]{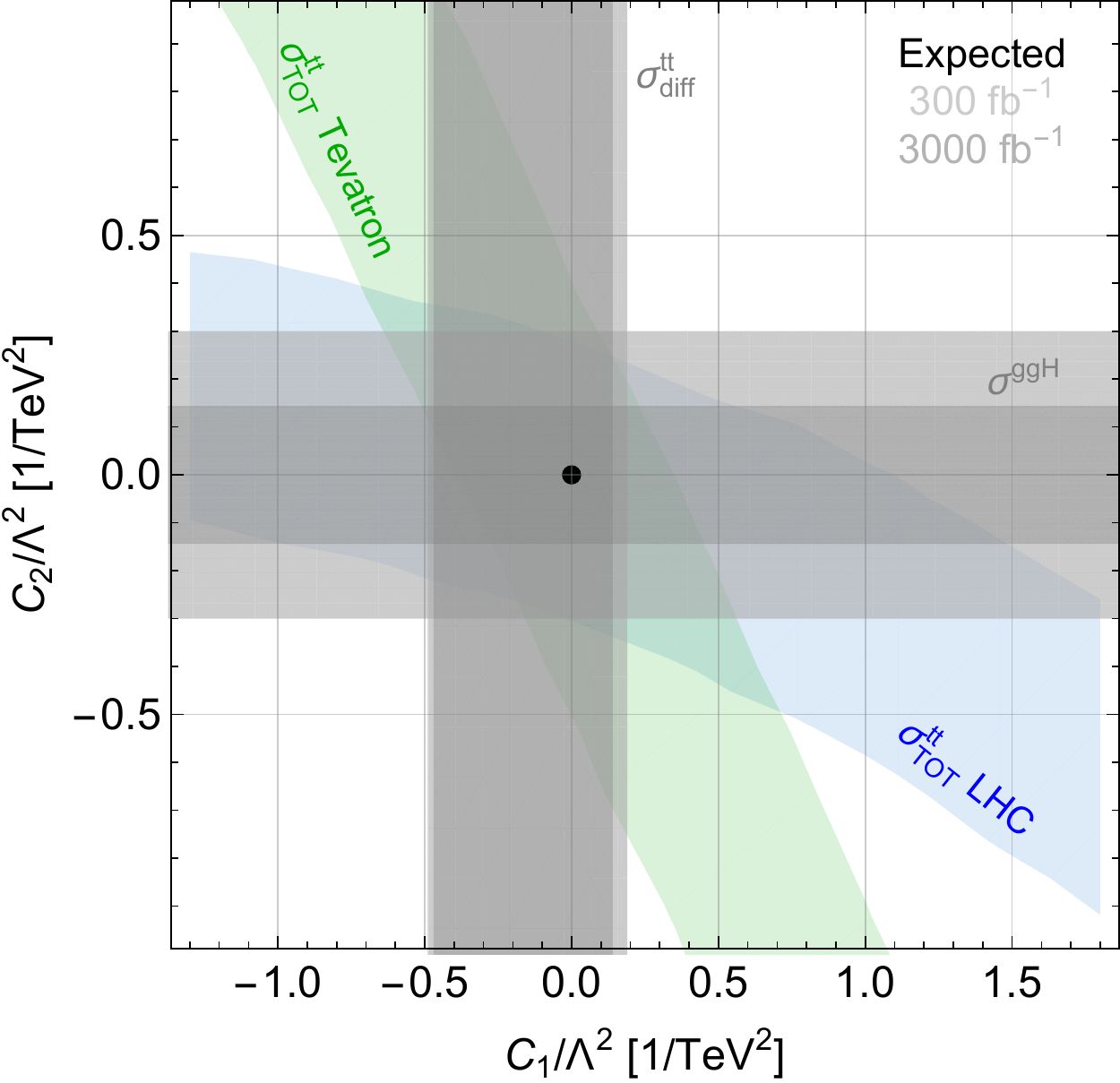}\\
\caption{95\% confidence intervals for  $C_1$ and $C_2$ from the measurements of the top quark transverse momentum differential cross section and Higgs production via gluon fusion cross section, vertical gray and horizontal gray shaded area respectively. The lighter (darker) gray area correspond to an integrated luminosity of 
300 (3000) fb$^{-1}$ respectively. The limits from the measurements of  $t\bar t$ total cross section at the LHC (blue) and Tevatron (green) are also shown.}
\label{fig:results_proj}
\end{center}
\end{figure}

Through this procedure we obtain the expected limits on the Wilson coefficient $C_1$ and $C_2$ shown in Fig.~\ref{fig:results_proj} where  light and dark gray regions correspond to an integrated luminosity of 300 and 3000 fb$^{-1}$ respectively. For comparison, the previous limits obtained from the measurements of the total $t\bar t$ cross section at Tevatron and LHC are also shown.  The plot shows that with an integrated luminosity of 300 fb$^{-1}$ the combination of the differential measurements in $t\bar t$ production together with the measurements of the Higgs production rate through gluon fusion will be able to set 
a comparable limits on the $C_2$ Wilson coefficient, and a stronger limit on $C_1$ for $C_1>0$. 
At the end of the LHC program, that is with a integrated luminosity of 3000 fb$^{-1}$, these measurements will provide the most stringent limits on the coefficient of the $\mathcal O_1$ and $\mathcal O_2$ operators. We report these values in Tab.~\ref{tab:future}.
All the limits can be turned around to be re-expressed as  lower bounds on $\Lambda$, the scale of the effective theory, by fixing $C_1=C_2= 4 \pi $ and taking the absolute value of the  limits in Table~\ref{tab:current}. Accordingly we find 
\be
\Lambda \gtap 4.3 \; \mbox{TeV}\quad \mbox{and} \quad \Lambda \gtap 5.5 \; \mbox{TeV}
\label{eq:lambda-limit}
\ee
from, respectively, the operator in Eq.~\eqref{q1} and Eq.~\eqref{q2}.  
The reliability of the expansion in the effective field theory approach is  verified if the probed energies $\bar E < \Lambda$. This is true for the Higgs production. It holds  for the differential top-pair production measurements analysis as well, even though in this case, as the explored transferred energies go up to about 3 TeV, we are approaching the limit.
The bounds of Eq.~\eqref{eq:lambda-limit} could be raised to almost 9 TeV with the expected reduced uncertainties.

\begin{table}[ht!]
\small
\begin{center}
\vspace{0.2cm}
\begin{tabular}{|c|c|}
\hline
 LHC 300 fb$^{-1}$   & LHC 3000 fb$^{-1}$  \cr
\hline
\hline
$\quad -0.49 < C_1/\Lambda^2 < 0.19 \quad$  & $ \quad -0.47 < C_1/\Lambda^2 < 0.19 \quad $  \cr
\hline
$\quad -0.30 < C_2/\Lambda^2 < 0.30 \quad$  & $ \quad -0.14 < C_2/\Lambda^2 < 0.14 \quad $ \cr
\hline
\end{tabular}
\end{center}
\caption{Expected limits at 95\% confidence level  on the coefficients  $C_{1}$ and $C_2$ from future data from  $ \di \sigma_{t \bar t}/\di p_{T}^t$ and $\mu_{ggH}$ respectively. Values are in units of TeV$^{-2}$.}
\label{tab:future}
\end{table}


\begin{acknowledgments}
We thank Marina Cobal and Michele Pinamonti for discussions. MF is associated to SISSA and the Department of Physics, University of Trieste.  The work of AT is supported by Coordena\c{c}\~ao de Aperfei\c{c}oamento de Pessoal de N\'ivel Superior (CAPES). AT would like to thank T. Hahn for the help with \texttt{FormCalc} and \texttt{LoopTools}. AT would like to thank ICTP-SIAFR and IFT-UNESP for hospitality.
\end{acknowledgments}
\appendix
\section{Relation with EFT}
\label{appendix}
Let us consider the operator ${\cal O}_{1}$ that we have introduced in \eq{q1} 
\begin{equation}
{\cal O}_{1}=\bar{t}\gamma^{\mu}T_{A}tD^{\nu}G_{\mu\nu}^{A}.
\end{equation}
It is possible to rewrite it as a specific combination of four-fermion operators belonging to the Warsaw basis~\cite{Grzadkowski:2010es}. In order to do that we perform an appropriate field redefinition by using the gluon equations of motion
\begin{equation}
D^{\nu}G_{\mu\nu}^{A}=-g_{s}\sum_{q}\bar{q}\gamma^{\mu}T^{A}q\,,
\end{equation}
where $\sum_q$ denotes the sum over all quarks. In this case we have
\begin{eqnarray}
{\cal O}_{1} & = & -g_{s}\bar{t}\gamma^{\mu}T_{A}t\sum_{q}\bar{q}\gamma^{\mu}T^{A}q\\
 & = & -g_{s}\bar{t}\gamma^{\mu}T_{A}t(\bar{u}\gamma^{\mu}T^{A}u+\bar{d}\gamma^{\mu}T^{A}d+\ldots)
\end{eqnarray}
where the ellipsis denote second and third generation quark currents. The relevant combination that enters in $t\bar{t}$ production at LHC and Tevatron is the one that couples the top-quark current with the
up- and down-quark current, namely
\begin{equation}\label{4fcomb}
\bar{t}\gamma^{\mu}T_{A}t(\bar{u}\gamma^{\mu}T^{A}u+\bar{d}\gamma^{\mu}T^{A}d).
\end{equation}
The following four-fermion operators of the Warsaw basis~\cite{Grzadkowski:2010es}  are those relevant for $t\bar{t}$ production induced by the partonic subprocesses $u\bar{u},d\bar{d}\to t\bar{t}$
\begin{eqnarray}\label{4fwars}
{\cal O}_{qq}^{(1)\,1331} & = & (\bar{u}_{L}\gamma_{\mu}t_{L})(\bar{t}_{L}\gamma^{\mu}u_{L})+\ldots\nn\\
{\cal O}{}_{uu}^{1331} & = & (\bar{u}_{R}\gamma_{\mu}t_{R})(\bar{t}_{R}\gamma^{\mu}u_{R})\nn\\
{\cal O}{}_{qq}^{(1)\,1133} & = & (\bar{u}_{L}\gamma_{\mu}u_{L})(\bar{t}_{L}\gamma^{\mu}t_{L})\nn\\
&&+(\bar{d}_{L}\gamma_{\mu}d_{L})(\bar{t}_{L}\gamma^{\mu}t_{L})+\ldots\nn\\
{\cal O}_{qq}^{(3)\,1133} & = & (\bar{u}_{L}\gamma_{\mu}u_{L})(\bar{t}_{L}\gamma^{\mu}t_{L})\nn\\
&&-(\bar{d}_{L}\gamma_{\mu}d_{L})(\bar{t}_{L}\gamma^{\mu}t_{L})+\ldots\nn\\
{\cal O}_{uu}^{1133} & = & (\bar{u}_{R}\gamma_{\mu}u_{R})(\bar{t}_{R}\gamma^{\mu}t_{R})\nn\\
{\cal O}{}_{qu}^{(8)\,1133} & = & (\bar{u}_{L}\gamma_{\mu}T^{A}u_{L})(\bar{t}_{R}\gamma^{\mu}T^{A}t_{R})\nn\\
&&+(\bar{d}_{L}\gamma_{\mu}T^{A}d_{L})(\bar{t}_{R}\gamma^{\mu}T^{A}t_{R})\\
{\cal O}{}_{qu}^{(8)\,3311} & = & (\bar{t}_{L}\gamma_{\mu}T^{A}t_{L})(\bar{u}_{R}\gamma^{\mu}T^{A}u_{R})+\ldots\nn\\
{\cal O}_{qq}^{(3)\,1331} & = & (\bar{u}_{L}\gamma_{\mu}t_{L})(\bar{t}_{L}\gamma^{\mu}u_{L})\nn\\
&&+2(\bar{d}_{L}\gamma_{\mu}t_{L})(\bar{t}_{L}\gamma^{\mu}d_{L})+\ldots\nn\\
{\cal O}{}_{ud}^{(8)\,3311} & = & (\bar{t}_{R}\gamma_{\mu}T^{A}t_{R})(\bar{d}_{R}\gamma^{\mu}T^{A}d_{R})\nn\\
{\cal O}{}_{qd}^{(8)\,3311} & = & (\bar{t}_{L}\gamma_{\mu}T^{A}t_{L})(\bar{d}_{R}\gamma^{\mu}T^{A}d_{R})+\ldots\nn\\\nn
\end{eqnarray}
where the ellipsis denote terms that do not contain two top quarks and two up quarks or two top quarks and two down quarks. In deriving  some of the expressions in Eq.~\eqref{4fwars} we made use of the Pauli matrices completeness relation
\begin{equation}
\sigma_{ij}^{I}\sigma_{kl}^{I}=2\delta_{il}\delta_{jk}-\delta_{ij}\delta_{kl}.
\end{equation}
By using the $SU(3)$ generators completeness relation
\begin{equation}
T_{ab}^{A}T_{cd}^{A}=\frac{1}{2}\delta_{ad}\delta_{bc}-\frac{1}{6}\delta_{ab}\delta_{cd}
\end{equation}
and the Fierz rearrangement for anticommuting spinors
\begin{equation}
\bar{\psi}_{1L}\gamma_{\mu}\psi_{2L}\bar{\psi}_{3L}\gamma^{\mu}\psi_{4L}=\bar{\psi}_{1L}
\gamma_{\mu}\psi_{4L}\bar{\psi}_{3L}\gamma^{\mu}\psi_{2L}
\end{equation}
\begin{equation}
\bar{\psi}_{1R}\gamma_{\mu}\psi_{2R}\bar{\psi}_{3R}\gamma^{\mu}\psi_{4R}=\bar{\psi}_{1R}
\gamma_{\mu}\psi_{4R}\bar{\psi}_{3R}\gamma^{\mu}\psi_{2R} \, .
\end{equation}
The operator ${\cal O}_{1}$ can be rewritten in terms of the following specific combination of four-fermion operators 
\begin{eqnarray}\label{lincomb}
\bar{t}\gamma^{\mu}T_{A}tD^{\nu}G_{\mu\nu}^{A}& = & \frac{1}{4}{\cal O}_{qq}^{(1)\,1331}-\frac{1}{6}{\cal O}{}_{qq}^{(1)\,1133}\nn\\
&&+\frac{1}{2}{\cal O}{}_{uu}^{1331}-\frac{1}{6}{\cal O}_{uu}^{1133}\nonumber \\
 &  & +{\cal O}{}_{qu}^{(8)\,1133}+{\cal O}{}_{qu}^{(8)\,3311}\nn\\
 &&+\frac{1}{4}{\cal O}{}_{qq}^{(3)\,1331} +{\cal O}{}_{ud}^{(8)\,3311}\nn\\
 &&+{\cal O}{}_{qd}^{(8)\,3311} \, . \nn
\end{eqnarray}
This operator equivalence holds when considering $t\bar{t}$ production
induced by the partonic subprocesses $u\bar{u},d\bar{d}\to t\bar{t}$. The operator ${\cal O}_{qq}^{(3)\,1133}$ does not enter in the linear combination of eq.~\eqref{lincomb}, and therefore any new physics that generates it is not captured by the operator ${\cal O}_{1}$.

More in general, in the EFT approach, the operators in eq.~\eqref{4fwars} enter in the $t\bar{t}$ production cross section induced by $u\bar{u},d\bar{d}$ in the initial state through four specific linear combinations of their coefficients~\cite{Zhang:2010dr,Buckley:2015nca}, namely
\bea 
C_u^1&=&6C_{qq}^{(1)1331}+3C_{uu}^{1331}\nn\\
&&-C_{qq}^{(1)1133}-C_{qq}^{(3)1133}-C_{uu}^{1133}\nn\\
C_u^2&=&-C_{qu}^{(8)1133}-C_{qu}^{(8)3311}\nn\\
C_d^1&=&3C_{qq}^{(3)1331}-3C_{qq}^{(1)1331}\nn\\
&&+C_{qq}^{(3)1133}-C_{qq}^{(1)1133}+6C_{ud}^{(8)3311}\nn\\
C_d^2&=&-C_{qu}^{(8)1133}-C_{qd}^{(8)3311} \, .
\eea
In case of ${\cal O}_{1}$ we have a unique coefficient and therefore the two approaches are not equivalent. 


\end{document}